\documentclass[12pt]{article}
\pdfoutput=1
\usepackage{amssymb}
\usepackage{amsmath}
\usepackage{amsfonts}
\usepackage{graphicx}
\usepackage{mathrsfs}
\usepackage{comment}
\usepackage{cite}

\begin{document}

\renewcommand{\figurename}{Fig.}
\renewcommand{\tablename}{Table.}
\newcommand{\Slash}[1]{{\ooalign{\hfil#1\hfil\crcr\raise.167ex\hbox{/}}}}
\newcommand{\bra}[1]{ \langle {#1} | }
\newcommand{\ket}[1]{ | {#1} \rangle }
\newcommand{\beq}{\begin{equation}}  \newcommand{\eeq}{\end{equation}}
\newcommand{\bef}{\begin{figure}}  \newcommand{\eef}{\end{figure}}
\newcommand{\bec}{\begin{center}}  \newcommand{\eec}{\end{center}}
\newcommand{\non}{\nonumber}  \newcommand{\eqn}[1]{\begin{equation} {#1}\end{equation}}
\newcommand{\laq}[1]{\label{eq:#1}}  
\newcommand{\dd}[1]{{d \o d{#1}}}
\newcommand{\Eq}[1]{Eq.(\ref{eq:#1})}
\newcommand{\Eqs}[1]{Eqs.(\ref{eq:#1})}
\newcommand{\eq}[1]{(\ref{eq:#1})}
\newcommand{\Sec}[1]{Sec.\ref{chap:#1}}
\newcommand{\ab}[1]{\left|{#1}\right|}
\newcommand{\vev}[1]{ \left\langle {#1} \right\rangle }
\newcommand{\bs}[1]{ {\boldsymbol {#1}} }
\newcommand{\lac}[1]{\label{chap:#1}}
\newcommand{\SU}[1]{{\rm SU{#1} } }
\newcommand{\SO}[1]{{\rm SO{#1}} }
\def\({\left(}
\def\){\right)}
\def\dt{{d \o dt}}
\def\diag{\mathop{\rm diag}\nolimits}
\def\Spin{\mathop{\rm Spin}}
\def\O{\mathcal{O}}
\def\U{\mathop{\rm U}}
\def\Sp{\mathop{\rm Sp}}
\def\SL{\mathop{\rm SL}}
\def\tr{\mathop{\rm tr}}
\def\ebq{\end{equation} \begin{equation}}
\newcommand{\OR}{~{\rm or}~}
\newcommand{\AND}{~{\rm and}~}
\newcommand{\EV}{ {\rm \, eV} }
\newcommand{\KEV}{ {\rm \, keV} }
\newcommand{\MEV}{ {\rm \, MeV} }
\newcommand{\GEV}{ {\rm \, GeV} }
\newcommand{\TEV}{ {\rm \, TeV} }
\def\o{\over}
\def\a{\alpha}
\def\b{\beta}
\def\c{\varepsilon}
\def\d{\delta}
\def\e{\epsilon}
\def\f{\phi}
\def\g{\gamma}
\def\h{\theta}
\def\k{\kappa}
\def\l{\lambda}
\def\m{\mu}
\def\n{\nu}
\def\p{\psi}
\def\q{\partial}
\def\r{\rho}
\def\s{\sigma}
\def\t{\tau}
\def\u{\upsilon}
\def\v{\varphi}
\def\w{\omega}
\def\x{\xi}
\def\y{\eta}
\def\z{\zeta}
\def\D{\Delta}
\def\G{\Gamma}
\def\H{\Theta}
\def\L{\Lambda}
\def\F{\Phi}
\def\P{\Psi}
\def\S{\Sigma}
\def\me{\mathrm e}
\def\ol{\overline}
\def\tl{\tilde}
\def\*{\dagger}

\newcommand{\rem}[1]{{$\spadesuit$\bf #1$\spadesuit$}}

%\begin{document}
%\begin{titlepage}
\begin{center}

\hfill KEK--TH--2136\\

\vspace{1cm}

{\Large\bf Explaining electron and muon $g-2$ anomaly \\ in SUSY without lepton-flavor mixings}
\vspace{1.5cm}

{\bf Motoi Endo$^{(a,b)}$ and Wen Yin$^{(c)}$}

{$^{\rm (a)}${\it
Theory Center, IPNS, KEK, Tsukuba, Ibaraki 305-0801, Japan}

\vskip 0.1in

$^{\rm (b)}${\it
The Graduate University of Advanced Studies (Sokendai),\\
Tsukuba, Ibaraki 305-0801, Japan}

\vskip 0.1in

$^{\rm (c)}${\it Department of Physics, KAIST, \\Daejeon 34141, Korea}
}

\vspace{12pt}
\vspace{1.5cm}

\date{\today $\vphantom{\bigg|_{\bigg|}^|}$}

\abstract{
We propose a SUSY scenario to explain the current electron and muon $g-2$ discrepancies without introducing lepton flavor mixings. 
Threshold corrections to the Yukawa couplings can enhance the electron $g-2$ and flip the sign of the SUSY contributions. 
The mechanism predicts a flavor-dependent slepton mass spectrum. 
We show that it is compatible with the Higgs mediation scenario.
 }

\end{center}
\clearpage

\setcounter{page}{1}
\setcounter{footnote}{0}

%\end{titlepage}
\setcounter{footnote}{0}
\section{Introduction}

%\section{Introduction}
The discrepancy of the lepton anomalous magnetic moment ($g-2$) is one of the leading candidates that indicate new physics beyond the standard model (SM). Both in the electron and muon sectors, the anomaly has been reported as
\begin{align}
\laq{g2e}
\D a_e &= a_e^{\rm EXP} - a_e^{\rm SM} = (-8.7 \pm 3.6) \times 10^{-13}, \\
\laq{g2m}
\D a_\mu &= a_\mu^{\rm EXP} - a_\mu^{\rm SM} = (27.4 \pm 7.3) \times 10^{-10},
\end{align}
where $a_\m^{\rm SM}$ is the SM prediction of the muon $g-2$~\cite{Davier:2017zfy,Keshavarzi:2018mgv}, and 
$a_\mu^{\rm EXP}$ is its experimental result~\cite{Bennett:2006fi,Roberts:2010cj}. 
Recently, a new discrepancy, $\D a_e$, was reported in the electron sector, due to the new measurement of the fine structure constant. See Refs.~\cite{Hanneke:2008tm,Hanneke:2010au} for the experimental value of the electron $g-2$, Ref.~\cite{Aoyama:2014sxa} for its theoretical prediction, and Ref.~\cite{Parker:2018vye} for the new result of the fine structure constant.

It is challenging to explain both anomalies theoretically. In a wide class of new physics models, contributions to the lepton $g-2$ are scaled by the lepton mass squared. Suppose the muon $g-2$ anomaly is a sign of new physics, the electron $g-2$ is expected to receive a contribution,
\begin{align}
\laq{ratio}
\frac{\D a_e}{\D a_\m} \sim \frac{m_e^2}{m_\m^2} \simeq 2.4\times 10^{-5}.
\end{align}
This is too small to explain the result \eq{g2e}. Thus, it seems to require very light new particles, which easily conflict with experimental constraints, e.g., from the LHC. In addition, the sign of Eq.~\eq{g2e} is opposite to Eq.~\eq{g2m}. Extra mechanisms may flip the sign. For instance, flavor violations in the lepton sector can solve the problems, though they are constrained tightly. 

New physics models have been studied to explain both anomalies~\cite{Davoudiasl:2018fbb,Crivellin:2018qmi,Liu:2018xkx,Dutta:2018fge,Han:2018znu}.
Within the context of the supersymmetry (SUSY), lepton flavor violations are examined~\cite{Dutta:2018fge}. SUSY contributions to the electron $g-2$ are enhanced by the tau Yukawa coupling via the mixings of the selectrons with the staus, instead of introducing very light SUSY particles. Further, the sign is chosen appropriately by the mixings. However, it was argued that the lepton-flavor violating $\tau \to e\g$ restricts the system. 

In this letter, we propose a new mechanism to explain both anomalies within the minimal supersymmetric standard model (MSSM). We assume the minimal flavor violation (MFV) for the lepton sector, and thus, the model is free from the lepton flavor violations. The key observation is threshold corrections to the lepton Yukawa couplings. They are non-linear in SUSY particle masses so that even if the SUSY particle masses follow the MFV hypothesis, the relation \eq{ratio} can be changed drastically. 
In particular, the SUSY electron Yukawa coupling can be enhanced, and its sign can be opposite to the muon one. The scenario predicts flavor-dependent slepton masses. We will discuss the Higgs mediation scenario as an explicit model~\cite{Yamaguchi:2016oqz}.

\section{Muon and electron $g-2$}
\lac{1}

The SUSY Yukawa couplings of leptons, $y_{i}$, are matched with the SM ones, $m_{i}/v$, non-trivially because of radiative corrections $\Delta_i$ and a ratio of the Higgs vacuum expectation values $\tan\b\equiv \left<H^0_u\right>/\left<H^0_d\right>$ as~\cite{Carena:1999py,Marchetti:2008hw,Hofer:2009xb}
\begin{align}
\laq{yukawas}
 y_{i} \simeq {m_{i}\o v}{ \sqrt{1+\tan^2\b} \o 1+\Delta_i},
\end{align}
where the Higgs vacuum expectation value is $v^2 = \left<H^0_u\right>^2 +\left<H^0_d\right>^2\simeq (174\GEV)^2$. In this Letter, we focus on a scenario with a large size of the Higgsino mass parameter, $\mu$, and large $\tan\b$. Then, the radiative corrections are dominated by threshold corrections from Bino-slepton loop diagrams. In the mass-insertion approximation, they become~\cite{Marchetti:2008hw}\footnote{
 In the numerical analysis, we use a general formula for evaluating $\Delta_i$, i.e., without assuming the mass-insertion approximation~\cite{Girrbach:2009uy}.
}
\begin{align}
 \Delta_i \simeq  \m \tan\beta \frac{g_Y^2 M_1}{16\pi^2} I(M_1^2, m_{\tl{ i}_L}^2, m_{\tl{i}_R}^2),
 \laq{delta}
\end{align}
with $i=e, \m$, and its superpartner $\tl{i}$. Here, $m_X$ is a mass of $X$, $g_Y$ is the gauge coupling of $\U(1)_Y$, and $M_1$ is the Bino mass. The loop function is defined as
\begin{align}
 I(x,y,z) = -\frac{xy\ln (x/y) + yz \ln (y/z) + zx \ln (z/x)}{(x-y)(y-z)(z-x)},
\end{align} 
which satisfies $I(x,x,x)=1/2x^2$.

By taking $m_{\tl{i}_L}=m_{\tl{i}_R}= M_1$, one obtains
\begin{align}
 \D_i \sim -1 \({\m \over -100\TEV }\) \({\tan\b\over 70}\)  \({ 2\TEV \over M_{1}}\).
\end{align}
When the Higgsino mass parameter is much larger than masses of the sleptons and the Bino, 
$|\Delta_i|$ can be as large as $\O(1)$. It is enhanced by $\m \tan\b$ coming from the trilinear coupling of $y_i \m H_u^\* \tl{i}_L\tl{i}_R. $\footnote{
 In general, there are also contributions from $y_i A_i  H_d \tl{i}_L \tl{i}_R$. We will omit this term for simplicity. The extension with it is straightforward. 
}
The sign of $\D_i$ can be either positive or negative depending on that of $\m$. When RG effects are neglected, $\Delta_i$ depends on a relative size of the soft breaking parameters and $\m$, and thus, does not change under the scaling, i.e., even by increasing the SUSY scale. 

Figure \ref{fig:del} shows $\D_i$ for varying the slepton soft masses with $m_{\tl{i}_L}=m_{\tl{i}_R}$, $M_1=1.5\TEV$, $M_2=500\GEV$, and $\tan\b=70$. Here, $\mu=-100\TEV$ (left) and $\mu=-500\TEV$ (right). 
The red and blue lines denote $\D_\m$ and $\D_e$, respectively.
It is found that $\D_i$ can be around or smaller than $-1$. In the discontinuity region of the red line, an eigenstate of the smuon becomes tachyonic.

The leading $\tan\beta$-enhanced radiative corrections are taken into account in Eq.~\eq{yukawas}, and $|\D_i|$ can be large~\cite{Marchetti:2008hw} (cf.~Ref.~\cite{Carena:1999py}).\footnote{
 Such a large $|\D_i|$ has been discussed in the context of the muon $g-2$~\cite{Borzumati:1999sp,Endo:2013lva,Bach:2015doa, Tran:2018kxv}.
}
They include a resummation of the radiative corrections in the form of $(g_Y^2\mu\tan\beta/M_{\rm SUSY})^n$ to all orders, where $M_{\rm SUSY}$ is a typical scale of SUSY particle masses in loops, while other corrections are suppressed. 

\begin{figure}[t!]
\begin{center}  
\includegraphics[width=70mm,bb=0 0 360 349]{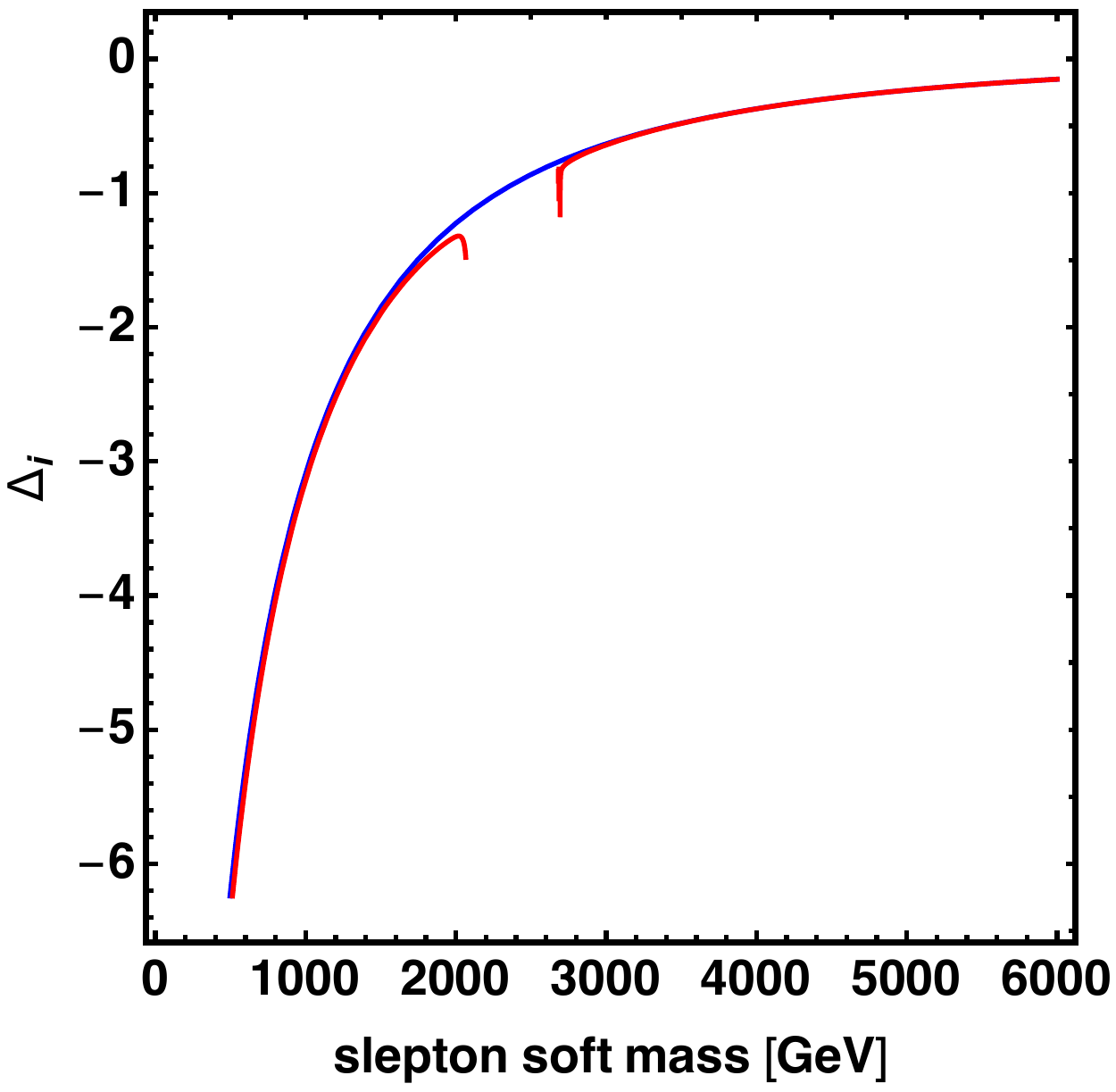}
\includegraphics[width=72mm,bb=0 0 360 340]{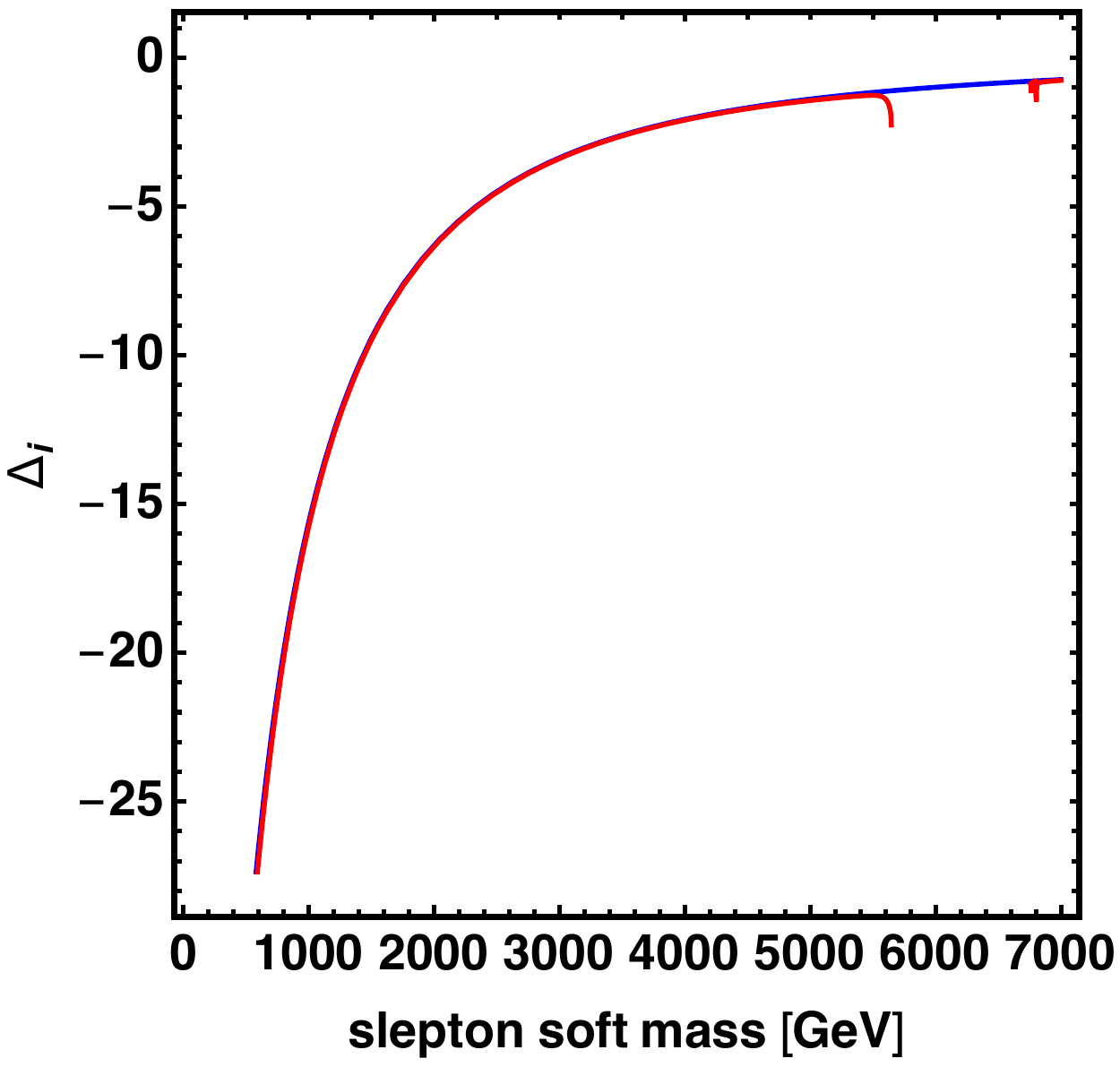}
\end{center}
\caption{$\D_e$ (blue) and $\D_\m$ (red) for varying the slepton soft mass. Here, $m_{\tl{i}_R}=m_{\tl{i}_L}, M_1=1.5\TEV, M_2=500\GEV,\tan\b=70,$ with $\mu=-100\TEV$ (left) and $\mu=-500\TEV$ (right). 
}
\label{fig:del}
\end{figure}

\begin{figure}[t!]
\begin{center}  
\includegraphics[width=74mm,bb=0 0 360 302]{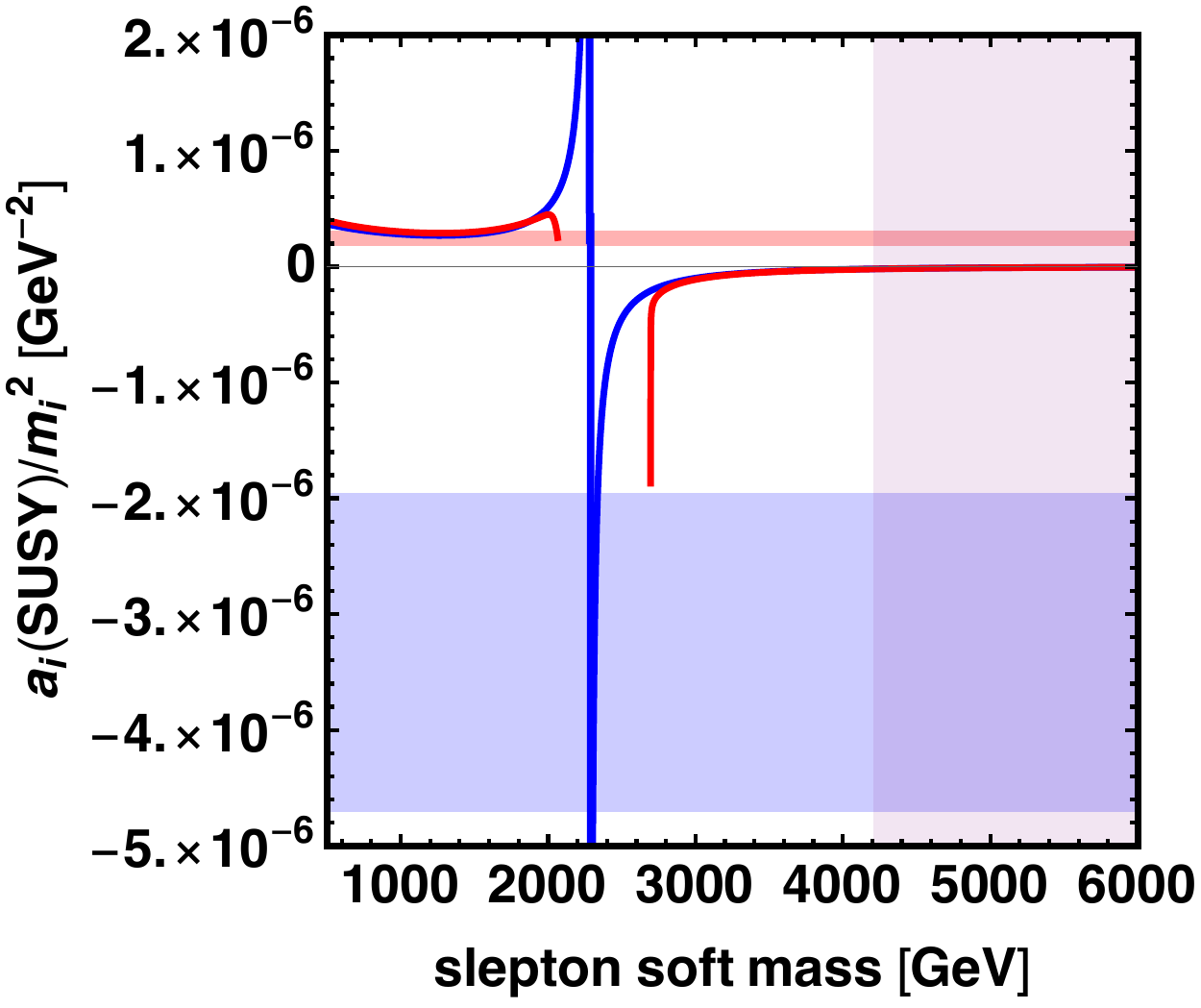}
\includegraphics[width=70mm,bb=0 0 360 319]{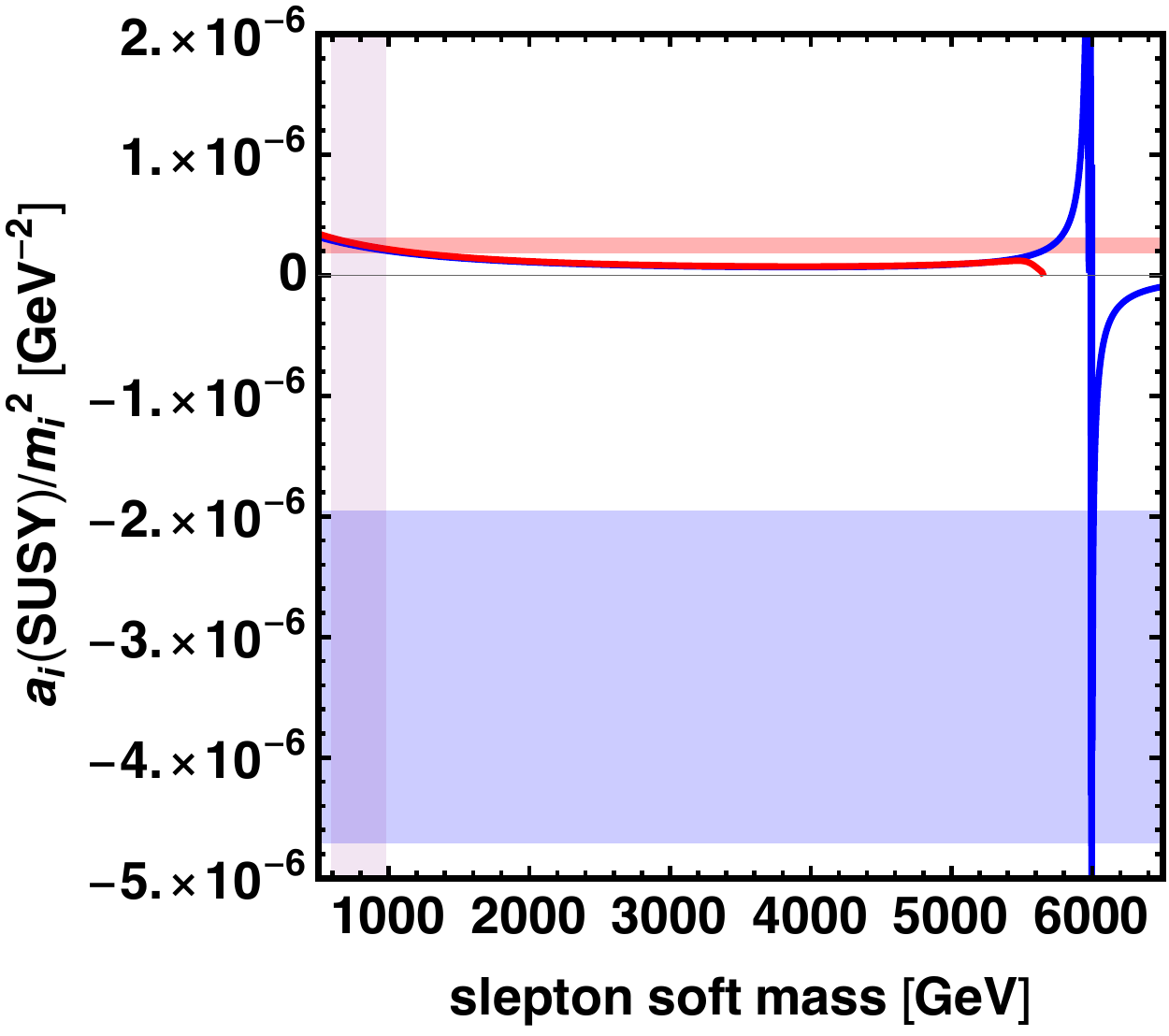}
\end{center}
\caption{$(a_e)_{\rm SUSY}/m^2_e$ (blue) and $(a_\m)_{\rm SUSY}/m^2_\m$ (red) for varying the slepton mass. The other parameters are same as Fig.\,\ref{fig:del}. In particular, $\mu=-100\TEV$ (left) and $\mu=-500\TEV$ (right). The light blue (red) horizontal band represents the observed discrepancy for the electron (muon) $g-2$ at the $1\s$ level. The smuons on the pink vertical band are stable against the vacuum decay at the tree level, where the model parameters are evaluated at the scale of the slepton soft mass.  } 
\label{fig:1}
\end{figure}

When $|\mu| \tan \b$ is large, the SUSY contributions to the lepton $g-2$ of the $i$-th generation, $(a_i)_{\rm SUSY}$, are dominated by the Bino-slepton diagrams.
In the mass-insertion approximation, they are represented as~\cite{Endo:2013lva}\footnote{
In the numerical analysis, we use the formula in Ref.~\cite{Lopez:1993vi,Chattopadhyay:1995ae,Moroi:1995yh} without the mass-insertion approximation for the one-loop contributions. In addition, the formula in Ref.~\cite{vonWeitershausen:2010zr} is used for $\delta_{\rm QED}$. 
}
\begin{align}
 (a_i)_{\rm SUSY} \simeq \left( \frac{1 - \delta_{\rm QED}}{1 + \Delta_i } \right) 
 {g_Y^2 \over 16\pi^2}{ m_i^2  \mu \tan\beta \, M_1 \over  m_{\tl{i}_L}^2 m_{\tilde{i}_R}^2}\,
 f_N\left( \frac{m_{\tilde{i}_L}^2}{M_1^2}, \frac{m_{\tilde{i}_R}^2}{M_1^2} \right), 
 \laq{gm2}
\end{align}
where $f_N(x,y)$ is the loop function defined in Ref.~\cite{Endo:2013lva} and satisfies $f_N(1,1)=1/6$. QED corrections beyond the leading order are taken into account by $\delta_{\rm QED}$~\cite{Degrassi:1998es}. The radiative correction $\D_i$ appears, because $(a_i)_{\rm SUSY}$ is proportional to the SUSY Yukawa coupling of the lepton. 

The SUSY contributions \eq{gm2} are scaled by the lepton mass squared $m_i^2$. It is noticed that $(a_i)_{\rm SUSY}$ can be affected drastically by $\D_i$, when $\mu M_1$ is large negative. 
For $\mu M_1 < 0$, $\D_i$ is negative.
Since $f_N(x,y)$ is positive, $(a_i)_{\rm SUSY}$ becomes positive (negative) for $1 + \D_i < 0\, (>0)$. In addition, \eq{gm2} is enhanced significantly around the cancellation point,
\begin{align}
 \Delta_i = -1.
\end{align}
Thus, $(a_i)_{\rm SUSY}/m^2_i$ can have different size and sign for different flavors, depending on $1 + \D_i$. It is noticed that lepton-flavor mixings are not necessary, and thus, there are no constraints from the lepton flavor violations. 

In Fig.~\ref{fig:1} we show $(a_i)_{\rm SUSY}/m_i^2$ for the electron (blue line) and the muon (red line).
In the horizontal blue band, the observed discrepancy for the electron $g-2$ (see Eq.~\eq{g2e}) is explained at the $1\s$ level, and that for the muon (see Eq.~\eq{g2m}) is shown by the red band. 
We find that the electron $g-2$ discrepancy is explained around the cancellation point $\D_e = -1$, corresponding to $m_{\tl{e}_{L,R}}\simeq 2.3 \TEV$ $(6.0\TEV)$ in the left (right) panel. The selectrons are relatively heavy and satisfy the collider constraints easily. On the other hand, the muon $g-2$ anomaly is explained by lighter smuons, because $(a_\mu)_{\rm SUSY}$ is required to be positive. Here, all the SUSY particles are set to satisfy the current collider/experimental bounds.\footnote{
 Light smuons can satisfy the LHC bounds, e.g., by setting the LSP appropriately.
}

Too large $|\mu| \tan \b$ spoils the stability of the electroweak vacuum. In the analysis, we used the formula provided in Ref.~\cite{Endo:2013lva} to derive the vacuum stability condition.\footnote{
 The formula fits the result of \texttt{CosmoTransitions 1.0.2}~\cite{Wainwright:2011kj} at the tree level. It may suffer from a large scale uncertainty~\cite{Endo:2015ixx}. In particular, an energy gap exists between the scales of the charge-color breaking vacuum, $\gtrsim 10^8\GEV$, and the electroweak vacuum, $\sim 100\GEV$. 
Since the potential can be lifted in a large renormalization scale, the constraint might be alleviated.
}
The trilinear coupling associated with $\mu \tan \b$ is proportional to the SUSY Yukawa coupling of the lepton. In the muon case, the vacuum is stable in the pink vertical band. There is a lower bound on the smuon masses, because the potential is stabilized when the smuons become heavy. In addition, an upper bound is obtained when the pink band appears to the left of the mass discontinuity region (see the right panel of Fig.~\ref{fig:1}). This is because, as the smuon masses increase, $|1 + \D_\m|$ decreases according to Fig.~\ref{fig:del}, and thus, the SUSY Yukawa coupling, i.e., the trilinear coupling of the smuons, is enhanced. In Fig.~\ref{fig:1}, it is found that the smuons are required to be heavier than $4.2\TEV$ for $\mu=-100\TEV$, while they are limited in $600\GEV \lesssim m_{\tl{\m}} \lesssim 1\TEV$ for $\mu=-500\TEV$ by the vacuum stability condition. In contrast, the vacuum stability constraint for the electron is highly alleviated and does not affect our scenario, because its Yukawa coupling is tiny.

Two sample points are given in Table.\,\ref{tab:1}. In both cases, the electron $g-2$ discrepancy is explained. For the muon, in order to satisfy the vacuum stability constraint, the smuon should be either heavier (left panel of Fig.~\ref{fig:1}) or lighter (right panel) than the selectron. In the latter case, the muon $g-2$ anomaly is explained with satisfying the vacuum stability constraint.\footnote{
The masses of stau, stop, and sbottom should also be large enough to avoid the vacuum stability bound, though they are irrelevant to the electron and muon $g-2$. A scenario satisfying this setup will be discussed in the next section. } Our sample points are consistent with the current LHC bounds.\footnote{
Large $\tan\b$ as much as $\gtrsim 70$ would not suffer from a Landau pole below the GUT scale $\sim 10^{16}\GEV$ if the gluino mass is large enough. The SUSY Yukawa coupling of the bottom quark can be suppressed by threshold corrections. 
} 
Consequently, we conclude that both the discrepancies of the electron and the muon $g-2$ can be explained simultaneously by choosing the slepton masses appropriately.

\begin{table*}[!t]
\begin{center}
\begin{tabular}{|c|c|c|}
\hline
& {\bf I} & {\bf II}   \\
\hline\hline
${\mu}$& $-100 $&  $-500$ \\
${\tan\b}$ & 70 & 70  \\
$M_1, M_2$ &1.5, 1.0 & 1.5, 0.6  \\
$m_{\tilde{e}_{L,R}}$  &2.4, 2.3 & 6.0, 6.0\\
$m_{\tilde{\m}_{L,R}}$ & 5.0, 5.0 & 0.7, 0.7  \\
\hline\hline
$\Delta_e$  &$-0.97$& $-0.99$ \\
$\Delta_\m$  &$-0.23$& $-23$ \\
$(a_{e})_{\rm SUSY}$ & $-8.8 \times 10^{-13}$ &  $-7.3 \times 10^{-13}$  \\
$(a_\mu)_{\rm SUSY}$ & $-0.1 \times 10^{-9}$ &  $3.1 \times 10^{-9}$  \\
\hline
\end{tabular}
\end{center}
\caption{Two sample points which explain the electron $g-2$ discrepancy. All masses are in units of TeV. The upper parameters are input, while the results are given below.}
\label{tab:1}
\end{table*}

Let us give three comments on the mechanism.
First of all, our analysis is almost independent of the Wino mass.
This is because the Wino diagrams relevant for $\D_i$ and $(a_i)_{\rm SUSY}$ internally exchange the Higgsino. The Higgsino is assumed to be so heavy that its contributions are suppressed. 
It is interesting to mention that our model can be compatible with the Wino LSP, which is a candidate of the dark matter. 

Secondly, let us mention how to test the mechanism. There are two ways, direct productions of the SUSY particles and indirect detections. The direct production of the selectrons are challenging, because they tend to be heavy for realizing $\D_e \simeq -1$. Their masses may exceed scopes of future collider experiments. In contrast, the smuons are as light as $\O(0.1-1)\TEV$, which could be tested in the LHC and future experiments. In particular, once the heavier smuon is produced, the branching fraction of its decay to the lighter one with the Higgs boson becomes sizable because of the large trilinear coupling. This may give a characteristic signature for the experiments. 

The scenario may be tested by indirect searches. 
Since $\D_i$ is close to $-1$ for the electron or large negative for the muon, the branching fractions of the (semi-) leptonic $B$ meson decays are affected when the heavy Higgs bosons are relatively light~\cite{Choudhury:1998ze,Babu:1999hn}.\footnote{
The quark sector also receives threshold corrections similarly. The SUSY Yukawa couplings of the down-type quarks can be enhanced with certain squark and gluino/Bino masses, and large $|\m| \tan\b$. Such effects may be observed in the quark flavor physics.  }
The decays can proceed by exchanging the heavy Higgs bosons, whose couplings to the leptons are given by the SUSY Yukawa couplings. The SUSY contributions to the muon channels are suppressed by large $|\D_\mu|^2$, whereas those to the electron modes are enhanced by $1/|1 + \D_e|^2$.
Next, SUSY corrections to the SM Higgs boson decaying into lepton pairs are weak (see Ref.~\cite{Endo:2015oia}). In fact, those to the muon channel are suppressed by large $|\D_\mu|^2$. For the electron channel, although the corrections are enhanced by $1/|1 + \D_e|^2$, they are still suppressed by $\cos(\beta-\alpha)$, where $\a$ is a Higgs mixing angle, and may be below sensitivities of the future electron-positron colliders. Further, if the Wino-like neutralino is the dark matter, the scenario can be tested from direct/indirect dark matter search experiments. 

Lastly, let us comment on a parameter tuning for $1 + \D_e = \O(0.1-1)\%$. 
This cancellation can be linked with the mass hierarchy between the electron and the muon, $m_e/m_\m=\O(0.1)\%$. 
The electron mass is realized by the SUSY Yukawa coupling, $y_e$, which is comparable to the muon one, 
because the Yukawa couplings satisfy the relation, 
\begin{align}
 {y_e \over y_\m} \simeq {m_e \over m_\m} {1+\Delta_\m \over 1+\Delta_e}.
\end{align}
In general, the small electron mass may be chosen by an anthropic selection~\cite{Agrawal:1997gf}.
Then, in our scenario, the selectron mass might be chosen to obtain the tiny electron mass.

\section{Higgs mediation scenario}

In order to explain the current discrepancies of the electron and muon $g-2$, the smuons are required to be lighter than the selectrons. In this section, we provide UV models to realize such a slepton spectrum. Let us assume the MFV for the slepton soft-breaking masses~\cite{Hall:1990ac, Ciuchini:1998xy, Buras:2000dm, DAmbrosio:2002vsn, Paradisi:2008qh},\footnote{
The smuons can also be embedded in $N=2$ SUSY multiplets~\cite{Shimizu:2015ara,Yin:2016pkz}. Here, SUSY breaking effects are suppressed due to the $N=2$ non-renormalization theorem. The smuons tend to be lighter than other sleptons.}
\begin{align}
m_{\tl{i}_L}^2 &\simeq d_L+c_L y_i^2, \notag \\
m_{\tl{i}_R}^2 &\simeq d_R+c_R y_i^2, \laq{model}
\end{align}
where higher order terms of $y_i$ are omitted. 
The first terms, $d_L$ and $d_R$, in the right-hand side are flavor-blind contributions, e.g., by SUSY-breaking mediations via gauge interactions.
The second terms, $c_L$ and $c_R$, depend on the lepton Yukawa couplings, i.e., depend on lepton flavors. Such contributions are yielded by SUSY-breaking mediations via the Higgs sector, as we will discuss below.
Here, the lepton Yukawa matrix is diagonalized without loss of generality, and hence, the slepton soft mass matrices are aligned to the Yukawa matrix. 
Then, there are no lepton-flavor violations.\footnote{
 This is not the case beyond the MSSM, e.g. when strongly-coupled right-handed neutrinos are introduced to explain the neutrino masses~\cite{Hisano:1995cp}. This may be supported by the thermal leptogenesis~\cite{Fukugita:1986hr}. Even in this case, one can introduce a flavor symmetry in the lepton Yukawa couplings, $y^N_i H_u L_i N_i$, where $N_i$ is the right-handed neutrinos. The neutrino oscillations are realized if the neutrino mass term, $W=\sum_{ij} M^N_{\rm ij} N_i N_j$, breaks the flavor symmetry. Flavor-violating effects from the neutrino Yukawa couplings should be suppressed. Also, the neutrino masses can be obtained by introducing the dimension five operator, $W= 1/M(L_j H_u)(L_i H_u)$, in a high energy scale. Then, the scenario does not change. The baryogenesis works with active neutrino oscillations when an inflaton decays to either the left-handed leptons flavor-dependently or the Higgs boson~\cite{Hamada:2018epb}. } 
In this Letter, we do not assume anything special for the squarks and the gluino. Their masses depend on details of the UV models. 

Let us discuss a Higgs mediation scenario to realize the flavor-dependent mass spectrum, $c_L$ and $c_R$. 
The scenario was first identified in a non-universal Higgs masses model~\cite{Yamaguchi:2016oqz}, where radiative corrections with negative large Higgs mass squares provide positive contributions to the squark and slepton masses which depend on the Yukawa couplings, i.e., flavors. 
By taking $m_{H_u}^2\simeq m_{H_d}^2 < 0$, the slepton masses are estimated by RG running as (cf., Ref.~\cite{Yanagida:2018eho}),
\begin{align}
\laq{Higmed}
 c_R\simeq  2c_L \simeq \frac{1}{4\pi^2} \bar{m}^2 \log{\(\frac{ M_{\rm GUT} }{\bar{m}}\)},
\end{align}
at the leading logarithmic approximation. Here, $M_{\rm GUT}\sim 10^{16}\GEV$ is the GUT scale, and $\bar{m}^2 \equiv -m_{H_d}^2$.
Tachyonic mass spectrum is avoided for the pseudo-Higgs boson by assuming~\cite{Yamaguchi:2016oqz, Yin:2016shg}
\begin{align}
 \m \sim -\bar{m},~\tan\b \gtrsim 50. 
\end{align}
This setup is favored to realize large $|\D_i|$. On the other hand, $d_L$ and $d_R$ depend on flavor-blind mediation mechanisms.\footnote{
The anomaly mediation has been discussed within the context of the Higgs mediation, which is called the Higgs-anomaly mediation~\cite{Yin:2016shg,Yanagida:2016kag,Yanagida:2018eho}. The anomaly mediation provides flavor-blind masses. 
Such a setup can be realized by sequestering sfermions and gauginos away from the SUSY breaking 
sector, while the two Higgs multiplets are not. Then, the soft mass parameters are vanishing for the sfermions and the gauginos at the input scale, but are not for the Higgs. The former masses are generated at loop levels.
Although this scenario can explain the muon $g-2$ anomaly, it is not possible to explain both the electron and muon $g-2$ anomalies simultaneously.
This is because the squarks in the first two generations become tachyonic when $\bar{m}$ becomes too large compared with the gaugino masses.
Such a difficulty is avoided if we take account of additional flavor-blind mediation. The additional contribution can induce large squark masses.}
In the following, we do not specify the mechanism and leave them as free parameters.

\begin{table*}[!t]
\begin{center}
\begin{tabular}{|c|c|c|c|}
\hline
 & {\bf I} & {\bf II}  & {\bf III}   \\
\hline\hline
$\bar{m}$& $100 $&  $350$ &  $350$ \\
${\tan\b}$ & 70 & 80  &  80\\
$\sqrt{d _L}$ & 2.0 & 0.28& 0.28  \\
$\sqrt{d_R}$ & 2.0 & 0.28& 0.28  \\
$M_1, M_2$ &1.0, 0.6 & 1.6, 0.6 & 1.6, 0.6 \\
\hline\hline
$m_{\tilde{e}_{1,2}}$  & 2.0, 2.0 & 6.4, 4.5& 6.4, 4.5\\
$m_{\tilde{\m}_{1,2}}$ & 4.9, 3.7 & 0.99, 0.64  & 16, 12\\
$m_{\tilde{\t}_{1,2}}$ & 56, 39 & 220, 150  & 220, 150\\
\hline\hline
$\Delta_e$  &$-0.96$& $-0.99$ & $-0.99$ \\
$\Delta_\m$  &$-0.23$& $-15$ & $-0.19$ \\
$|y_e|$  &$0.005$& $0.023$ & $0.023$ \\
$|y_\m|$  &$0.06$& $0.003$ & $0.062$ \\
$(a_{e})_{\rm SUSY} $& $-6.9 \times 10^{-13} $ &  $-5.6  \times 10^{-13} $  &  $-5.6  \times 10^{-13}$ \\
$(a_\m)_{\rm SUSY}$& $-0.2 \times 10^{-9}$ &  $2.6 \times 10^{-9}$ &  $-0.01 \times 10^{-9}$  \\
\hline
\end{tabular}
\end{center}
\caption{Higgs mediation sample points. All masses are in units of TeV. The model input parameters are provided above, and the results are given in middle and below. The selectron, smuon, and stau masses are shown in the middle. }
\label{tab:2}
\end{table*}

There are two types of mass spectra for smuons and selectrons which are consistent with the Higgs mediation. 
According to the previous section, when $m_{\tl{\mu}}\gg m_{\tl{e}}$, one obtains $|\D_\mu|\ll1$ and  $|y_\m|\gtrsim |y_e|$. On the other hand, when $m_{\tl{\mu}}\ll m_{\tl{e}}$, $|\D_\mu|$ can be so large that $|y_\m|$ becomes smaller than $|y_e|$. These two spectra can be realized by the Higgs mediation. 
In fact, when $\bar{m}^2>0$ they satisfy the following relation,
\begin{align}
\laq{Higgs}
 \(\frac{m^2_{\tl{\m}_{L,R}}-m^2_{\tl{e}_{L,R}}}{y_\mu^2-y_e^2}\)>0.
\end{align}

Let us provide three data points of the Higgs mediation scenario in Table \ref{tab:2}. 
The model parameter $\bar{m}$ is input at the GUT scale, and the flavor-dependent contributions to the slepton masses are derived by solving the RG equations, i.e., by using Eq.~\eq{Higmed}. On the other hand, the flavor-blind contributions, $d_L$ and $d_R$, as well as the gaugino masses, $M_1$ and $M_2$, are free parameters in our analysis. Their values at the scale $\bar{m}$ are also provided in Table \ref{tab:2}. Then, the soft masses and the SUSY Yukawa couplings are derived by using Eq.~\eq{model} with the threshold corrections, $\Delta_i$. 
Points {\bf I} and {\bf III} explain the electron $g-2$ discrepancy, while the muon $g-2$ anomaly is not. On the other hand, both are explained at Point {\bf II}. In all cases, the vacuum stability condition is satisfied for the stau as well as the smuon, because the staus become heavy in the scenario. 
It is noticed that Points $\bf II$ and $\bf III$ have the same dimensionful input parameters, despite that the results are different. This is because multiple sets of the smuon SUSY-breaking masses satisfy Eq.~\eq{yukawas}. Then, the dimensionless parameters, particularly $y_\m$, become different due to large threshold corrections.

Before closing this section, let us mention about the stop and sbottom masses. They are likely to be as large as $\O(10-100)\TEV$ by the Higgs mediation similarly to the stau. Such a setup can be consistent with the SM Higgs boson mass and the vacuum stability condition in the squark sector particularly by assuming the gluino mass appropriately~\cite{Vega:2015fna}. However, we face with a severe little hierarchy problem due to the large stop masses. The discussion on this problem is beyond the scope of this letter and will be studied elsewhere.

\section{Conclusions}

We proposed an MSSM scenario to explain both the electron and muon $g-2$ discrepancies without introducing lepton flavor mixings. The discrepancies are different in scale and sign. The electron $g-2$ requires larger SUSY contributions than the muon $g-2$ with an opposite sign. In our scenario, this is realized by the threshold corrections to the SUSY Yukawa interactions with the flavor-dependent slepton mass spectrum. The electron Yukawa coupling becomes enhanced by them, and its sign can be opposed to the muon one. In order to explain both anomalies, the smuons are required to be (much) lighter than the selectrons. We discussed that such a mass spectrum is consistent with the Higgs mediation scenario.

\vspace{1em}
\noindent {\it Acknowledgements}: \\
W.Y. thanks the hospitality of the KEK theory center where this work was initiated.
This work was supported by JSPS KAKENHI No.~16K17681 (M.E.) and 16H03991 (M.E.), and NRF Strategic Research Program NRF-2017R1E1A1A01072736 (W.Y.).

\providecommand{\href}[2]{#2}\begingroup\raggedright\endgroup

\end{document}